\documentclass[12pt,preprint]{aastex}
\usepackage{graphicx}

\newcommand{\X}{\mathcal{X}}
\newcommand{\Y}{\mathcal{Y}}
\newcommand{\D}{\mathcal{D}}

\newcommand{\bfx}{{\bf x}}
\newcommand{\bfy}{{\bf y}}
\newcommand{\bfz}{{\bf z}}

\newcommand{\bfb}{{\bf b}}
\newcommand{\bfk}{{\bf k}}

\begin{document}

\title{Novel Methods for Predicting Photometric Redshifts from
Broad Band Photometry using Virtual Sensors}

\author{M. J. Way}
\affil{NASA Ames Research Center, Space Sciences Division,
MS 245-6, Moffett Field, CA 94035, USA}
\author{A. N. Srivastava}
\affil{NASA Ames Research Center, Intelligent Systems Division,
MS 269-4, Moffett Field, CA 94035, USA}

%\date{Received December 28, 2004 / Accepted 2005}

\begin{abstract}
We calculate photometric redshifts from the Sloan Digital Sky Survey
Main Galaxy Sample, The Galaxy Evolution Explorer All Sky Survey,
and The Two Micron All Sky Survey using two new training--set
methods. We utilize the broadband photometry from the three surveys
alongside Sloan Digital Sky Survey measures of photometric quality
and galaxy morphology. Our first training--set method draws from the
theory of ensemble learning while the second employs Gaussian
process regression both of which allow for the estimation of
redshift along with a measure of uncertainty in the estimation. The
Gaussian process models the data very effectively with small
training samples of approximately 1000 points or less.  These two
methods are compared to a well known Artificial Neural Network
training--set method and to simple linear and quadratic regression. 
Our results show that robust photometric redshift errors as low as
0.02 RMS can regularly be obtained.  We also demonstrate the need
to provide confidence bands on the error estimation made by both
classes of models.  Our results indicate
that variations due to the optimization procedure used for almost
all neural networks, combined with the variations due to the data
sample, can produce models with variations in accuracy that span an
order of magnitude. A key contribution of this paper is to quantify
the variability in the quality of results as a function of model and
training sample.  We show how simply choosing the ``best" model
given a data set and model class can produce misleading results.

\end{abstract}

\keywords{Photometric Redshifts, Sloan Digital Sky Survey, Galaxy Evolution
Explorer All Sky Survey, Two Micron All Sky Survey}

%%%%%%%%%%%%%%%%%%%%%%%%%%%%%%%%%%%%%%%%%%%%%%%%%%%%%%%%%%%%%%%%%%%%%%%

\section{INTRODUCTION}

Using broadband photometry in multiple filters to estimate
redshifts of galaxies was likely first attempted by \citet{Baum62}
on 25 galaxies in nine broadband imaging filters in the visible and
near--infrared range. Given the low throughput of spectrographs much
is to be gained by attempting to estimate galaxy redshifts from
broadband colors rather than from measurement of individual
spectra. In the Sloan Digital Sky Survey \citep[SDSS,][]{York2000}
100 million galaxies will have accurate broadband u,g,r,i,z photometry,
but only 1 million galaxy redshifts from this sample will be
measured. If a method can be found to obtain an accurate estimate of
the redshift for the larger SDSS photometric catalog, rather than
the smaller spectroscopic one, much better constraints on the
formation and evolution of large--scale structural elements such as
galaxy clusters, filaments, and walls and cosmological models
in general \citep[e.g.][]{BlakeBridle2005} may be achieved.

Two approaches, spectral energy distribution fitting (SED fitting:
also known as ``template--fitting") and the training--set method (TS method),
have been used to obtain photometric redshifts over the past 30 years.
In order to use TS methods galaxies with a similar range in magnitude and
color over the same possible redshift range must be used to estimate the
redshifts from the broadband colors measured. Since this type of data has
not always been available SED fitting has historically been the preferred method
\citep[e.g.][]{Koo85,Loh86,Lanzetta96,Kodama99,Benitez00,Massarotti01,Babbedge04,Padmanabhan05}
given the historically low numbers of galaxies with spectroscopically confirmed
redshifts in deep photometric surveys of the universe. This is due to the fact
that photometric surveys have always gone, and continue to go, deeper than is
possible with spectroscopy.  Another alternative has been to use training sets
consisting of a combination of both observed galaxy templates and those from
galaxy evolution models \citep[e.g. hyperz,][]{Bolzonella00}.

There are many approaches to SED fitting. For example, \citet{Kodama99}
use four--filter ({/it BVRI}) photometry and a Bayesian classifier
using SED fitting which they have tested out to z=1 and claim is valid beyond
this redshift.  The approach of \citet{Benitez00} makes use of additional
information such as the shape of the redshift distributions and fractions of
different galaxy types. This may be helpful in instances where one has
a limited sample size at large redshifts. However, all estimators,
Bayesian or otherwise, can be biased due to small sample size effects.

TS methods rely on having a complete sample of galaxies in magnitude,
color and redshift. Hence these methods have been restricted to
relatively nearby z$<$1 surveys, such as the SDSS, rather than much
deeper surveys such as the Hubble Deep Field \citep{Williams96}.
In fact, for redshifts above 1 there have not been sufficiently large
and complete enough measured samples of galaxy redshifts, magnitudes and
colors to use TS methods with much accuracy \citep[e.g.][]{Wang98}.
As well, the colors of galaxies change nearly monotonically up to z$=$1,
but beyond this the color--redshift space becomes much more complex and simple linear
and quadratic regression will fail. Hence SED fitting has been used almost
exclusively for surveys of z$>$1.  See \citet{Benitez00} for an excellent
detailed discussion of the differences and similarities between these two
commonly used approaches.

In the past 10 years a large number of empirical fitting techniques
for TS methods have come into use and new techniques continue to be
developed. Some examples of linear and non--linear methods include:
2nd and 3rd order polynomial fitting \citep{Brunner97,Wang98,Budavari05};
quadratic polynomial fitting \citep{Hsieh05,Connolly95};
support vector machines \citep{Wadadekar05}; nearest neighbor and kd--trees
\citep{Csabai03}, and artificial neural networks
\citep{Firth03,Tagliaferri03,Ball04,Collister04,Vanzella04}.

\begin{figure}[tbp]
\plottwo{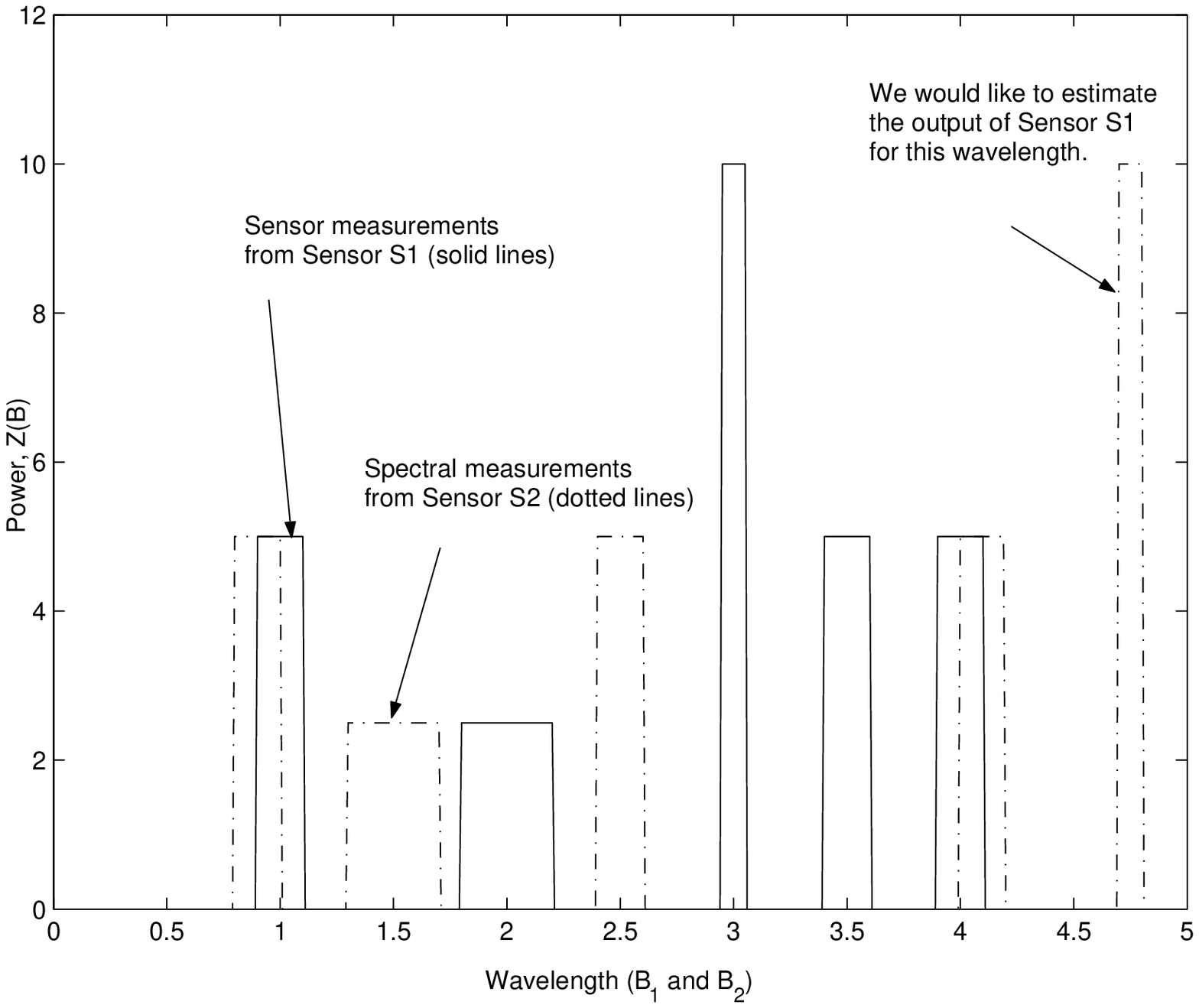}{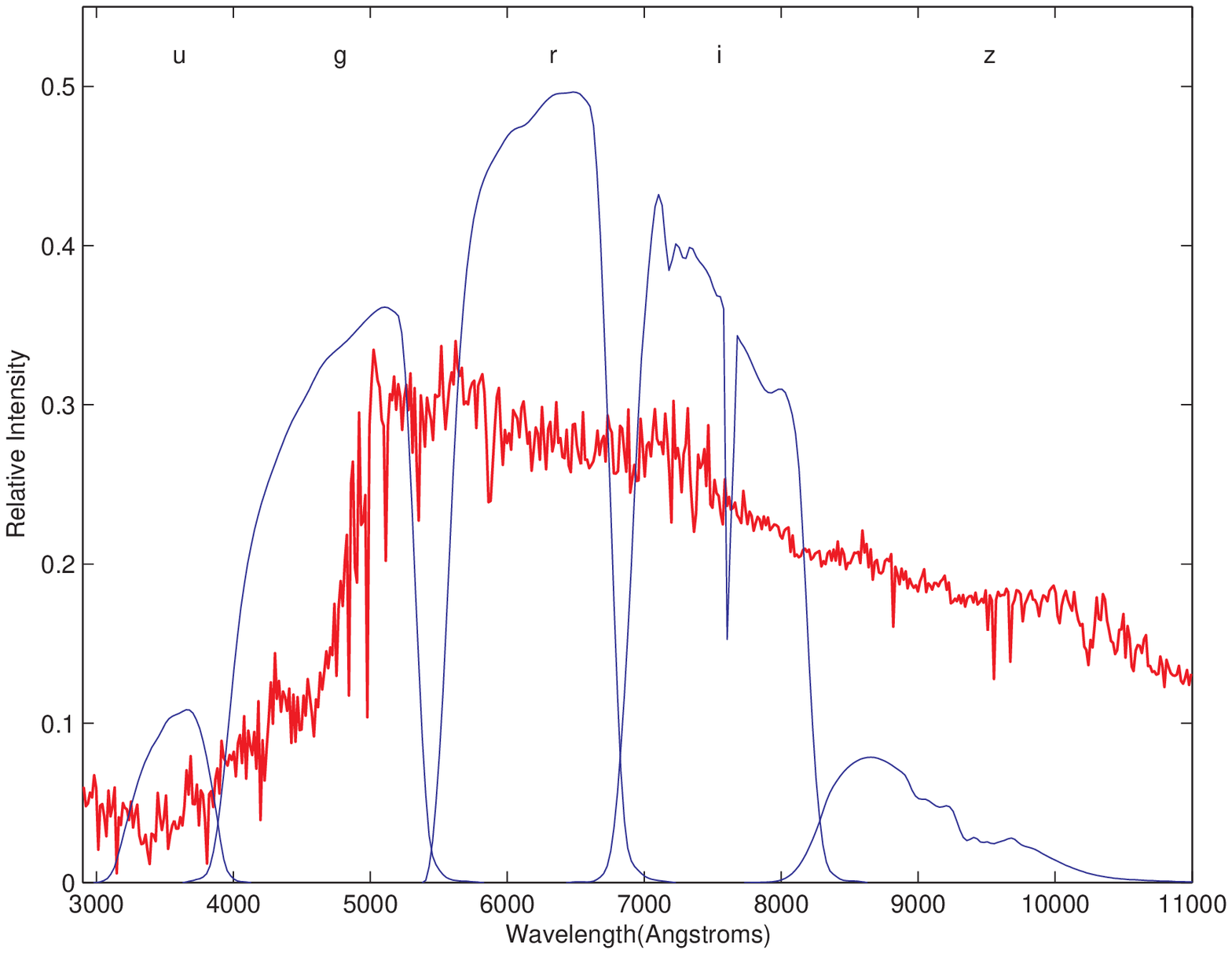}
\caption{The left figure is a cartoon to help illustrate the need for a Virtual
Sensor.  We have spectral measurements from two sensors $\mathcal{S}_1$ and
$\mathcal{S}_2$, (solid and dot--dashed lines, respectively). We wish to estimate the
output of sensor $\mathcal{S}_1$ for a wavelength where there is no
actual measurement from the sensor.  Note that some sensor
measurements overlap perfectly, as in the case of wavelength $= 3$,
and in other cases, such as wavelength = 1, there is some overlap in
the measurements. The right figure shows the sensitivity through an airmass
of 1.3 for extended sources in the five SDSS (u,g,r,i,z)
filter bandpasses with the spectrum of NGC5102 \citep{Bergmann95}
purposely redshifted 1000{\AA} overlayed.}
\label{fig:figure1}
\vspace*{-.2in}
\end{figure}

We explore the problem of estimating redshifts from broadband
photometric measurements using the idea of a virtual sensor
\citep{SrivastavaOza2005,SrivastavaStroeve2003}.  These methods
allow for the estimation of unmeasured spectral phenomena based on
learning the potentially nonlinear correlations between observed
sets of spectral measurements. In the case of estimating redshifts,
we can learn the nonlinear correlation between spectroscopically
measured redshifts and broadband colors.  Statistically speaking,
this amounts to building a regression model to estimate the
photometric redshift.  However, the procedure is much more
complex than a simple regression due to the  significant effort
required for model building and validation.  The concept of
virtual sensors applies to the entire chain of analytical steps
leading up to the prediction of the redshift.
Figure~\ref{fig:figure1} shows a schematic of the
assumptions behind a Virtual Sensor with a cartoon on the left and
the real--world case with the five SDSS bandpasses and a sample
galaxy spectrum overlaid on the right.

As a baseline comparison, results from a TS--based neural network
package called ANNz \citep{Collister04} are presented.  Linear and
quadratic fits along the lines discussed in \cite{Connolly95} are
also presented.  Unlike all other previous work, we also discuss the
application of bootstrap resampling \citep{Efron79,ET93} for the
linear, quadratic, and ANNz models.

We apply the TS methods discussed above to the SDSS five--color
({\it ugriz}) imaging survey known as the Main Galaxy Sample
\citep[MGS,][]{Strauss02} which has a large calibration set of
spectroscopic redshifts for the SDSS Data Release 2
\citep[DR2,][]{Abazajian04} and SDSS Data Release 3
\citep[DR3,][]{Abazajian05}. The Two Micron All Sky
Survey \citep[2MASS,][]{Skrutskie06}\footnote{http://www.ipac.caltech.edu/2mass/}
extended source catalog along with Galaxy Evolution Explorer
\citep[GALEX,][]{Martin05}\footnote{http://www.galex.caltech.edu/}
data are also used in conjunction with the SDSS
where all three overlap to create a combined catalog for use with
our TS methods.

The data sets used in our analysis are discussed in {\S} 2,
discussion of the photometry and spectroscopic quality of the data sets
along with other photometric pipeline output properties of interest is given
in {\S} 3, the classification schemes used to obtain photometric redshifts are
in {\S} 4, comparison of the results takes place in {\S} 5, and we
summarize in {\S} 6.

\begin{table}
\begin{center}
\caption{Survey filters and characteristics \label{tbl-1}}
\begin{tabular}{ccrrc}
\hline \hline
Bandpass& Survey& $\lambda_{eff}$ & $\Delta\lambda$ & FWHM \tablenotemark{1}\\
        &       &    ({\AA})      &     ({\AA})     &   ($\arcsec$)         \\
\hline
FUV     & GALEX & 1528       & 442       & 4.5 \\
NUV     & GALEX & 2271       & 1060      & 6.0 \\
u       & SDSS  & 3551       & 600       & 1-2 \\
g       & SDSS  & 4686       & 1400      & 1-2 \\
r       & SDSS  & 6165       & 1400      & 1-2 \\
i       & SDSS  & 7481       & 1500      & 1-2 \\
z       & SDSS  & 8931       & 1200      & 1-2 \\
j       & 2MASS & 12500      & 1620      & 2-3 \\
h       & 2MASS & 16500      & 2510      & 2-3 \\
k$_{s}$ & 2MASS & 21700      & 2620      & 2-3 \\
\hline \hline
\end{tabular}
\tablenotetext{1}{The Full Width at Half Maximum is dependent on the seeing at
the time of the observation for ground based data.}
\end{center}
\end{table}

\section{THE SLOAN DIGITAL SKY SURVEY, THE TWO MICRON ALL SKY SURVEY and
THE GALAXY EVOLUTION EXPLORER DATA SETS}

Most of the work herein is related to the SDSS MGS DR2 and DR3, and
the photometric quantities associated with them. For completeness we
have added the 2MASS extended source catalog and GALEX
All Sky Survey photometric attributes where data exists for the same SDSS
MGS galaxies with corresponding redshifts. The 2MASS and GALEX data samples
are small where they overlap with those of the SDSS MGS galaxies with
corresponding known spectroscopic redshifts in the DR2 and DR3.
However, they appear copious enough for our new TS methods as there is
no evidence of over--fitting of these smaller data samples.

The Sloan Digital Sky Survey \citep{York2000} will eventually
encompass roughly 1/4 of the entire sky, collecting five--band
photometric data in 7700 deg$^{2}$ down to 23rd magnitude in r of
order 10$^{8}$ celestial objects. For about 1 in every 100 of these
objects down to g$\sim$20 a spectrum will be measured, coming to a
total of about 10$^{6}$ galaxy and quasar redshifts over roughly the
same area of the sky (7000 deg$^{2}$) as the photometric survey
\citep{Stoughton02}. The five broadband filters used, u,g,r,i and
z, cover the optical range of the spectrum (Table~\ref{tbl-1}).

We use several catalogs derived from the SDSS.  The MGS \citep{Strauss02}
of the SDSS is a magnitude--limited survey that targets all galaxies down
to r$_{Petrosian}$$<$17.77.  We use the MGS from DR2 and DR3 where spectroscopic
redshifts exist in order to validate our methods.

The 2MASS extended source catalog contains positions and magnitudes in j, h,
and k$_{s}$ filters for 1,647,599 galaxies and other nebulae across the entire
sky (Table~\ref{tbl-1}). The extended source magnitude limits in the three
filters are j=15.0, h=14.3 and k$_{s}$=13.5.  See \cite{Jarrett00}
for more detailed information on the extended source catalog.

The GALEX data release 1 (GR1)\footnote{http://galex.stsci.edu/GR1/}
All Sky photometry catalog contains positions and magnitudes in two
ultraviolet bands called the far ultraviolet Band (FUV) and the near
ultraviolet band (NUV).  See Table~\ref{tbl-1} for details on these broadband
pass filters. Limiting magnitudes for the all--sky (100 s integrations)
FUV is 19.9, and 20.8 for the NUV. See
\cite{Morrissey05} and references therein for more details of the
in--orbit instrument performance and \cite{Martin05} for mission
details.  The all--sky GR1 covers 2792 deg$^{2}$ of the sky.

\section{PHOTOMETRIC AND REDSHIFT QUALITY, MORPHOLOGICAL INDICATORS, AND
OTHER CATALOG PROPERTIES}

Historically most determinations of photometric redshifts from
large photometric surveys contain only broadband magnitudes
without reference to other parameters that may have been available from the
photometric aperture reductions themselves.  With the SDSS most papers
have utilized only the five band photometry ({\it ugriz}) while a host
of additional parameters like Petrosian radii \citep{Strauss02}, measures
of ellipticity \citep{Stoughton02}, and other derived quantities are
readily available from the photometric pipeline reductions.

This section explains the various quality flags used to obtain data
from the SDSS photometric and redshift catalogs, the photometric
catalogs of the 2MASS extended source catalog, and the GALEX All Sky
Survey.  We also explore the mophological indicators most likely to yield
information related to the prediction of redshifts in the SDSS MGS
for our TS calculations. 
The last subsection ({\S} 3.6) describes the four data set types used
in our analysis.

\subsection{The SDSS photometric quality flags}

The SDSS photometric pipeline \citep{Lupton01} produces a host of quality flags
\citep[Table 9]{Stoughton02} giving additional
information on how the photometry was estimated. The
primtarget flag is used to make sure the MGS is chosen
and extinction--corrected model magnitudes \citep{Stoughton02}
are used throughout this work (see query in Appendix I).

Herein we define GOOD and GREAT quality photometry
(see Table~\ref{tbl-2} for a description) where ! means NOT:

\noindent GOOD: !BRIGHT and !BLENDED and !SATURATED

\noindent GREAT: GOOD and !CHILD and !COSMICRAY and !INTERP

\begin{table}
\scriptsize
\begin{center}
\caption{Photometric Quality Flags used in this paper \tablenotemark{a}
\label{tbl-2}}
\begin{tabular}{ccl}
\hline \hline
Name& Bitmask  & Description \\
\hline
BRIGHT   & 0x00002&Object detected in first bright object finding step; generally brighter than r=17.5\\
BLENDED  & 0x00008&Object had multiple peaks detected within it\\
SATURATED& 0x40000&Object contains one or more saturated pixels\\
CHILD    & 0x00010&Object product of attempt to deblend BLENDED object\\
COSMICRAY& 0x01000&Contains pixel interpreted to be part of a cosmic ray\\
INTERP   & 0x20000&Object contains pixel(s) values determined by interpolation\\
\hline \hline
\end{tabular}
\tablenotetext{a}{\cite{Stoughton02}}
\end{center}
\end{table}

In this manner one can determine whether a difference in the quality of the
photometry makes any difference in the errors of the estimated photometric
redshifts. The only reason not to always use the very best photometry
(what we call GREAT in this work) is that the total number of
galaxies can drop by orders of magnitude and hence one may end up
sampling a much smaller number of objects. However, not
everyones needs are the same and hence the quality can be weighted based on
what is desirable. See Appendix I for the complete SDSS skyserver\footnote{
http://casjobs.sdss.org} queries used to obtain the data used in this paper.

\subsection{The SDSS redshift quality flags}

The SDSS spectroscopic survey \citep{Stoughton02,Newman04} has
several flags to warn the user of poor--quality redshifts that come
from the spectroscopic pipeline reductions \citep{Stoughton02}. This
is important because an inaccurate training set will result in poor
results no matter which method is used. To this end we utilized an
estimate of the confidence of the spectroscopic redshift called
zConf. Hence only those galaxies with zConf$>$0.95 in the MGS are
chosen. Other authors \citep[e.g.][]{Wadadekar05} have chosen to
use only the zWarning flag set to zero.  Our studies find zConf
values far below that of 0.95 when only the zWarning$=$0 flag is
set. This may put into question the reliability of such redshift
estimates. In addition, by setting zConf to values greater than
0.95, as we have done, the zWarning$=$0 flag is also included.
Extensive color--color, color--magnitude and magnitude error
plots were checked against galaxies with values of zConf$<=$0.95
and those with zConf$>$0.95. No clustering was found in any
of these plots related to zConf values and hence no color or magnitude
bias is introduced by the exclusion of zConf$<=$0.95 data.

\subsection{2MASS photometric quality and cross--reference with the SDSS}

Given the high quality constraints of the published photometry of the
2MASS extended source public release catalog \citep{Jarrett00},
only one quality flag is
checked. The extended source catalog confusion flag, ``cc\_flg",
is required to be zero in all three band passes.

The j\_m\_k20fe, h\_m\_k20fe, and k\_m\_k20fe isophotal fiducial elliptical
aperture magnitudes as defined in the 2MASS database are extracted for
the respectively described j, h, and k$_{s}$ 2MASS magnitudes used in this work.

The extended source catalog was loaded into our local SQL database containing
the SDSS DR2 to create a combined catalog (see next section).

\subsection{GALEX photometric quality and cross--reference with the SDSS}

Near--ultraviolet (nuv) and far--ultraviolet (fuv) broadband photometry are
extracted from the GALEX database for our use.  Several quality flags are
used to make sure the data are of the highest quality. Bad
photometry values in nuv photometry (nuv\_mag) and fuv
photometry (fuv\_mag) are given the value of --99 in the GR1 database,
and these are excluded from our catalog if either or both filters
contain such a value. The nuv\_artifact=0 flag is set to avoid
all objects with known bad photometry artifacts. Hence if
nuv\_artifact has any value other than zero the nuv\_mag is
considered bad.  Currently fuv\_artifact is always zero in the GR1.
The band$=$3 flag is used since it indicates detection in both nuv
and fuv bands. Finally, a value of fov\_radius$<$0.55 is required as this is
the minimum recommended value to make sure the distance of the object in degrees
from the center of the field of view of the telescope is not too
large, as this is known to cause problems in the quality of the
photometry obtained.

As with the 2MASS extended source catalog, the GALEX All Sky Survey data
were loaded into our local SQL database now containing the SDSS DR2
and 2MASS catalogs. The SDSS MGS with redshifts and the 2MASS
extended source catalogs were cross--referenced with GALEX when all
three catalog positions agreed to within 5$\arcsec$. The methods and
results used are comparable to those of \cite{Seibert05}: hence we
do not go further into a description of the combined catalog. See
Appendix I for a sample query.

\subsection{SDSS Petrosian Radii, Inverse Concentration Index, FracDev,
and Stokes}

The photometry properties discussed below are available in all five
SDSS bandpasses ({\it ugriz}), but we use the r--bandpass values for these quantities
as, in general, the r--band result has the lowest error and gives more
consistent results. This is also reasonable given the low redshifts
used, but this strategy would be questionable at higher redshifts
when morphological features in the rest frame r band start to get
more strongly shifted to the i and z bands.

It has been shown that using \citet{Petrosian76} 50\% and 90\% flux
radii \citep[e.g.][]{Wadadekar05} in addition to the SDSS five--band
photometry one can improve results by as much as 15\% (see Table~\ref{tbl-3}).
The Petrosian 50\% (90\%)
radius is the radius where 50\% (90\%) of the flux of the object is contained.
Given the low redshifts of this catalog they can be assumed to be a rough
measure of the angular size of the object.
The ratio of these quantities is called the Petrosian inverse concentration
index (CI) 1/c $\equiv$ r$_{50}$/r$_{90}$ which measures the slope of the
light profile.  The concentration index corresponds nicely to eyeball
morphological classifications of large nearby galaxies
\citep{Strateva01,Shimasaku01}.

The Petrosian Radii are also used in combination with a measure of the
profile type from the SDSS photometric pipeline reduction called
FracDev. FracDev comes from a linear combination of the best
exponential and de Vaucouleurs profiles that are fit to the image
in each band. FracDev is the de Vaucouleurs term \citep[\S 3.1,
][]{Abazajian04}. It is 1 for a pure de Vaucouleurs profile
typical of early--type galaxies and zero for a pure exponential
profile typical of late--type galaxies. FracDev is represented as a
floating point number between zero and 1.  This is similar to the
use of the S\'{e}rsic n--index \citep{Sersic68} for morphological
classification. The idea of using FracDev as a proxy for the
S\'{e}rsic index n comes from \citet{Vincent04} who show that if
S\'{e}rsic profiles with 1$<$n$<$4 accurately describe the SDSS
galaxy early and late types then FracDev is a ``monotonically
increasing function of the S\'{e}rsic index n, and thus can be used
as a surrogate for n.'' For a recent discussion on S\'{e}rsic
profiles see \citet{Graham05}. \citet{Blanton03a,Blanton03b} have
also shown that S\'{e}rsic fits to the azimuthally averaged radial
profile of an SDSS object provide a better estimate of galaxy
morphology than the Petrosian inverse concentration index
(1/c$\equiv$r$_{50}$/r$_{90}$) for the majority of MGS objects.
However, at the time of this work these profiles were only available
in the derived SDSS DR2 NYU-VAGC catalog of \citet{Blanton05}, and
our own studies do not show appreciable improvement over the
Petrosian inverse concentration index when used to calculate
photometric redshifts.

Measures of galaxy ellipticity and orientation, as projected on the sky, can
be obtained from the SDSS photometric pipeline ``Stokes'' parameters Q and U
\citep{Stoughton02}. These are the flux--weighted second moments of
a particular isophote.
\begin{equation} M_{xx}\equiv\langle\frac{x^{2}}{r^{2}}\rangle, \ \ \
M_{yy}\equiv\langle\frac{y^{2}}{r^{2}}\rangle, \ \ \
M_{xy}\equiv\langle\frac{xy}{r^{2}}\rangle
\end{equation}
According to \citet{Stoughton02} when the isophotes are self--similar ellipses
one finds
\begin{equation}
Q\equiv M_{xx}-M_{yy}=
\frac{a-b}{a+b}\cos(2\phi), \ \ \
U\equiv M_{xy}=\frac{a-b}{a+b}\sin(2\phi),
\end{equation}

Since the Stokes values are related to the axis ratio and position
angle, using these quantities in combination with those above should
give additional information on the galaxy types we are sampling and
hence help in the estimation of photometric redshifts. However, in
our studies we only utilize the Q parameter defined above as we see
no improvement when using both Q and U.

\subsection{Description of the four data set types used}

Four classes of data sets are used in our analysis, based on the
descriptions above.

\noindent \underline{Data set 1:} SDSS MGS GOOD quality photometry. All of the
data come from the SDSS MGS with the GOOD quality flags set. There are six
subsets in this data set as seen in Figure~\ref{fig:figure3}.
\begin{enumerate}
\item u-g-r-i-z: contains only the SDSS five--band extinction corrected magnitudes.
\item u-g-r-i-z-petro50-petro90: contains the u-g-r-i-z data and the Petrosian
        50\% and 90\% radii in the r band.
\item u-g-r-i-z-petro50-petro90-ci: contains the u-g-r-i-z-petro50-petro90
        data and the Petrosian concentration index as described in {\S} 3.5.
\item u-g-r-i-z-petro50-petro90-ci-qr: contains the
         u-g-r-i-z-petro50-petro90-ci and the Stokes Q parameter
        as described in {\S} 3.5.
\item u-g-r-i-z-petro50-petro90-fracdev: contains the
         u-g-r-i-z-petro50-petro90 and the FracDev parameter
        as described in {\S} 3.5.
\item u-g-r-i-z-petro50-petro90-qr-fracdev: contains the
         u-g-r-i-z-petro50-petro90-fracdev and the Stokes Q parameter
        as described in {\S} 3.5.
\end{enumerate}
    Each subset consists of 202,297 galaxies.

\noindent \underline{Data set 2:} SDSS MGS GREAT quality photometry. All of the
data, as seen in Figure~\ref{fig:figure4}, come
from the SDSS MGS with the GREAT quality flags set. There are six subsets
named and described in the same way as for data set 1.
Each subset consists of 33,328 galaxies.

\noindent \underline{Data set 3:} GALEX GR1, SDSS MGS GOOD quality photometry,
and the 2MASS extended source catalogs labeled as nuv-fuv-ugriz-jhk.
As seen in Figure~\ref{fig:figure5}a,
it consists of the two ultraviolet magnitudes from the GALEX GR1 database
(nuv and fuv).  It has the five SDSS MGS extinction--corrected magnitudes
(u,g,r,i,z) with the GOOD quality photometry flags set, but unlike
data sets 1 and 2 there are no other SDSS inputs used.  It also
contains the three 2MASS extended source catalog magnitudes
(j,h,k$_{s}$).  The total data set consists of 3095 galaxies.

\noindent \underline{Data set 4:} GALEX GR1, SDSS MGS GREAT quality photometry,
and the 2MASS extended source catalogs. As shown in
Figure~\ref{fig:figure5}b it is nearly the
same as data set 3, except the SDSS MGS GREAT quality photometry
flags are set.  The total data set consists of 326 galaxies.

\section{TRAINING METHODS}

We estimate the photometric redshifts of the galaxies in the SDSS, 2MASS
and GALEX databases using several classes of algorithms: simple linear and
quadratic regression, neural networks, and Gaussian processes.  These methods
have different properties and make different assumptions about the underlying
data generating process that will be discussed below.

\subsection{Linear and Quadratic fits}

Linear and quadratic polynomial fitting along the lines of
\cite{Connolly95,Hsieh05} are used as a way to benchmark the new
methods discussed below.  The linear regression for the SDSS {\it ugriz}
magnitudes would be given by an equation of the form:

\begin{equation}
Z=A+Bu+Cg+Dr+Ei+Fz
\end{equation}

Where A, B, C, D, E, and F result from the fit. All
data points are weighted equally. Z is the redshift: the spectoscopic
one when training and the photometric one when testing.

The quadratic form is similar and again all points are
weighted equally.

\begin{equation}
\scriptsize
Z=A+Bu+Cg+Dr+Ei+Fz+Guu+Hgg+Irr+Jii+Kzz+Lug+Mur+Nui+Ouz+Pgr+Qgi+Rgz+Sri+Trz+Uiz
\end{equation}

\subsection{The Artificial Neural Network approach}

The artificial neural network (ANNz) approach of \cite{Collister04}
is specifically designed to calculate photometric redshifts from any
galaxy properties the user deems desirable.  It has been demonstrated
to work remarkably well on the SDSS DR1 \citep{Collister04}. The ANNz
package contains code to run back--propagation neural networks with arbitrary
numbers of hidden units, layers and transfer functions. We chose two hidden
units, and 10 nodes in each of these units
(see Figure~\ref{fig:figure2}). See the next section for
a more detailed description of neural networks in general, or see
\cite{Collister04}.

\subsection{The Ensemble Model}
Back--propagation neural networks have been used extensively in a
variety of applications since their inception.  A good summary of
the methods we use can be found in \citet{bish95}.  Neural networks
are a form of nonlinear regression in which a mapping, defined as a
linear combination of nonlinear functions of the inputs, are used to
approximate a desired target value.  The weights of the linear
combination are usually set using an approach, based on gradient
descent of a cost function, that is defined between the target value
and the estimated value.  The cost function usually has multiple
local minima, and the model obtained at the end of a training cycle
usually corresponds to one such minima and not to a global minimum.
The global minimum would correspond to the model that best
approximates the training set.  Generalization of the model on a
test set (i.e., data that is not used during the model building
phase) can be shown to be poor if a global minimum is reached due to
the phenomenon of over--fitting.

The following material is a standard demonstration that although the
neural network computes a nonlinear function of the inputs,
distribution of errors follows a Gaussian if the squared error cost
function is minimized.  The cost function encodes an underlying
model of the distribution of errors. For example, suppose we are
given a data set of inputs $\X$, targets $\Y$, and a model
parameterized by $\Theta$. The standard method of obtaining the
parameter $\Theta$ is by maximizing the likelihood of observing the
data $\D = (\X, \Y)$ with the model $\Theta$.  Thus, we need to
maximize:
\begin{eqnarray}
\nonumber  P(\Theta | \D) &=& \frac{P(\D | \Theta)P(\Theta)}{P(\D)}\\
\nonumber  &\propto& P(\D|\Theta)P(\Theta)
\end{eqnarray}
and we note that $P(\D|\Theta) = P(\X, \Y | \Theta)$ and so:
\begin{equation}
\nonumber  P(\X, \Y | \Theta) = P(\Y | \X,\Theta)P(\X|\Theta)
\end{equation}
The function $P(\Theta)$ represents the prior distribution over
model parameters.  If we have knowledge about the ways in which the
weights of the model are distributed \emph{before the data arrives},
such information can be encoded in the prior. \cite{Neal1996} has
shown that in the limit of an infinitely large network, certain
simple assumptions on the distribution of the initial weights make a
neural network converge to a Gaussian process.  If we assume that
the errors are normally distributed, we can write the likelihood of
an input pattern $\bfx_i \in \X$ having target $y_i \in \Y$ with
variance $\sigma^2$ as~\footnote{We follow the convention that
bold--faced notation indicates vectors and non--bold faced symbols
indicate scalars}:
\begin{eqnarray}
\nonumber  L(y_i | \bfx_i, \Theta) &=& P(y_i|\bfx_i, \Theta) \\
\nonumber  &=& \frac{1}{\sqrt{2 \pi} \sigma} \exp -\frac{\left( y_i
- \hat{y}_i\right)^2}{2 \sigma^2}
\end{eqnarray}
The product of these likelihoods across the $N$ data points in the
data set $\D$ is the likelihood of the entire data set:
\begin{eqnarray}
\nonumber  P(\Y | \X, \Theta) &=& \prod_{i=1}^{N} P(y_i|\bfx_i,
\Theta) \\
  &=& \prod_{i=1}^{N} \frac{1}{\sqrt{2 \pi} \sigma} \exp
-\frac{\left( y_i - \hat{y}_i\right)^2}{2 \sigma^2}
\end{eqnarray}
From this equation, it is straightforward to see that maximizing the
log of this likelihood function is equivalent to minimizing the
squared error, which is the standard cost function for feed--forward
neural networks used in regression problems.

\begin{figure}[tbp]
\centering
\includegraphics[width=2.5in]{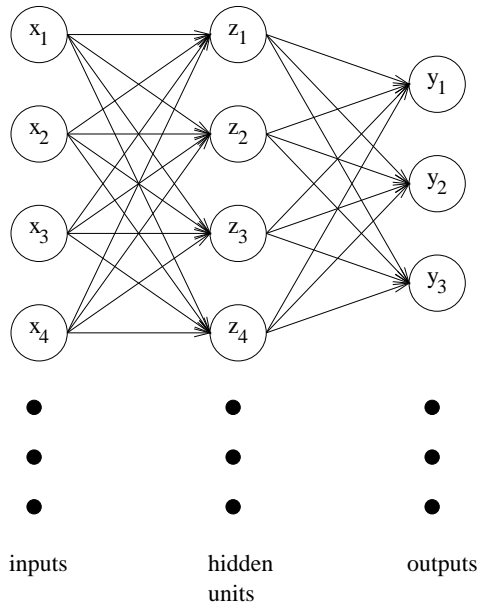}
\caption{A graphical depiction of a neural network with 4 inputs, 4
hidden units, and 3 outputs.  The outputs are nonlinear functions of
the inputs.} \label{fig:figure2} \vspace*{-.2in}
\end{figure}

Neural networks are often depicted as a directed graph consisting of
nodes and arcs as shown in Figure~\ref{fig:figure2}. For a
$p$ dimensional input $\bfx$ the value at the $k$ hidden nodes
$\bfz$ is the $k \times 1$ vector:
\begin{equation}
\bfz = s(W_1\bfx + \bfb_1)
\end{equation}
and the final estimate of the target $y$ is given by $\hat{y}$:
\begin{eqnarray}
\nonumber \hat{y} &=& W_2\bfz + \bfb_2 \\
                  &=& f(\bfx, \Theta)
\end{eqnarray}
where $W_1$ is a $k \times p$ matrix, $\bfb_1$ is a $p \times 1$
vector, $W_2$ is a $k \times l$ matrix and $\bfb_2$ is an $l \times
1$ vector.  In the case where the network only generates one output
per input pattern as is the case in the studies presented here,
$l=1$.

The function $s$ is a nonlinear function and is chosen as a
sigmoid:
\begin{eqnarray}
s(a) \equiv \frac{1}{1+exp(-a)}.
\end{eqnarray}

Neural networks are trained to fit data by maximizing the likelihood
of the data given the parameters.  The model obtained through this
maximization process corresponds to a single model sampled from the
space of models parameterized by the model parameters $\Theta.$  If
we assume Gaussian errors, we have shown that the cost function is
the well--known sum--squared error criterion. The network is trained
by performing gradient descent in the parameter space $\Theta.$ The
derivative of this cost function with respect to each weight in the
network is calculated and the weights are adjusted to reduce the
error. Because the cost function is non--convex, the optimization
problem gets caught in local minima, thus making training and model
optimization difficult.
In order to reduce the effects of local minima, we performed
\emph{bagging} or Bootstrap AGgregation \citep{breiman96bagging}. In
this procedure, we sample the data set $\D$ $M$ times with
replacement.  For each sample, we build one neural network in the
ensemble of $M$ neural networks.  The final prediction is formed by
taking the mean prediction of all $M$ neural networks:
\begin{equation}
\hat{y} = \frac{1}{M} \sum_{i=1}^M \hat{y_i}
\end{equation}
\citet{breiman96bagging} showed that this procedure results in a
regression model with lower error.  Our results, which we term our
``ensemble model'' (see Tables \ref{tbl-4}--\ref{tbl-6}),
show the effects of the local minima and the distribution of errors that
result from this problem on the SDSS, 2MASS, and GALEX data sets.

\subsection{Kernel Methods and Gaussian Processes}
In many ways, neural networks are attractive models for nonlinear
regression problems because they can scale to large data sets, and
provide a good baseline from which to compare other methods.  In the
machine learning literature, kernel methods have in many ways
subsumed neural networks because it was shown that as the number of
hidden units increases, if we assume that the weights and biases of
the neural network are drawn from a Gaussian distribution (thus
assuming that $P(\Theta)$ is Gaussian), the prior distribution over
functions implied by such weights and biases converges to a Gaussian
process \citep{Neal1996,CristianiniShawe2000}.

To describe a Gaussian process, we first note that in the case of a
neural network, $\hat{y}$ is defined as a specific nonlinear
function of $\bfx$, parametrized by $\Theta$, $\hat{y} = f(\bfx,
\Theta)$. In a Gaussian process, we actually define a prior
distribution over the space of functions $f$ which is assumed to be
Gaussian.  Thus, we have:
\begin{equation}
P[(f(\bfx_1), f(\bfx_2), ..., f(\bfx_N)) = (y_1, y_2, ..., y_N)]
\propto \exp(- \frac{1}{2} \bfy^T \Sigma^{-1}\bfy ).
\end{equation}
The marginals for all subsets of variables of a Gaussian process are
Gaussian.  The covariance matrix $\Sigma$ measures the degree of
correlation between inputs $\bfx_i$ and $\bfx_j$.  The choice of the
correlation function $\Sigma$ defines a potentially nonlinear
relationship between the inputs and the outputs.  If we choose
$\Sigma(\bfx_i, \bfx_j) = K(\bfx_i, \bfx_j)$, where $K$ is a
positive definite function, we obtain a specific Gaussian process
induced by the kernel function $K$.  To make a prediction with a
Gaussian process, we assume that a covariance function has been
chosen, and then compute:
\begin{equation}
P(y_{N+1}) = \frac{P(y_{N+1}, \bfy)}{P(\bfy)}
\end{equation}
We know that this distribution will be Gaussian, and the mean and
variance of the distribution can be computed as
follows \citep{CristianiniShawe2000}:

\begin{eqnarray}
\hat{y}_{N+1} = f(\bfx_{N+1}) &=& \bfy^T (K + \lambda^2 I)\bfk \\
\sigma^2(\bfx_{N+1}) &=& \Sigma(\bfx_{N+1}, \bfx_{N+1}) - \bfk^T(K +
\lambda^2I)^{-1}\bfk
\end{eqnarray}
where $\bfk = \Sigma(\bfx_i, \bfx)$, $K = K(\bfx_i, \bfx_j)$, and
$\lambda$ is an externally tuned parameter that represents the noise
in the output.

The nonlinearity in the model comes from the choice of the kernel
function $K$.  Typical choices for $K$ include the radial basis
function: $K(\bfx_i, \bfx_j) = \exp( - \frac{1}{2\sigma^2}||\bfx_i -
\bfx_j||^2 ) $ or the polynomial kernel $K(\bfx_i, \bfx_j) = (1 +
\bfx_i^T\bfx_j)^r$. We choose the latter for this study.

It can be shown that Gaussian process regression, as described
above, builds a linear model in a very high dimensional feature
space that is induced by the nonlinear kernel function $K$.
One distinct advantage of the Gaussian process is that it delivers
point predictions as well as a confidence interval around the
predictions. 

\section{DISCUSSION}

Results discussed below include the two different SDSS photometric quality
flag combinations used called GOOD and GREAT. For the SDSS data 10 different
photometric pipeline output parameters are utilized in different combinations
(see {\S} 3.6): u,g,r,i, and z extinction corrected model magnitudes, r band
Petrosian 50\% flux radii (petro50) and Petrosian 90\% flux radii (petro90),
the Petrosian inverse concentration index (CI) derived from these two
quantities, the r band FracDev quantity (FD), and r band Stokes value all
as defined in {\S} 3.5 and 3.6. Results are also discussed from the combined
catalogs of the SDSS MGS (u,g,r,i,z magnitudes only) galaxies with redshifts,
the 2MASS extended source catalog (j,h,k$_{s}$ magnitudes), and the GALEX
All Sky Survey (nuv,fuv magnitudes) data sets.  The sample sizes for each of
these data sets are also given in Tables~\ref{tbl-4}--\ref{tbl-6}.

In order to make our results as comparable as possible the same
validation, training and testing sizes are used in our analysis for
ANNz, ensemble model, linear, and quadratic fits: training=89\%,
validation=1\%, and testing=10\%.  In order to put proper confidence intervals
on the error estimates from these methods, bootstrap
resampling \citep{Efron79,ET93} is utilized on the training data:
90\% of the training data are used for each of 100 bootstraps.

For the Gaussian processes the situation is slightly different.
The same percentages for training, validation, and testing are utilized.
However, for data sets 1-3 1000 samples from the training data
are used for each of the bootstrap runs. For data set 4 only 50 samples
are utilized for each of the bootstrap runs.
The Gaussian processes require matrix inversion which is an
$O(N^{3})$ operation. Hence small training sets were required to
complete this project in a reasonable time frame.

\begin{figure}[dr3dresults]
\epsscale{0.8}
\plotone{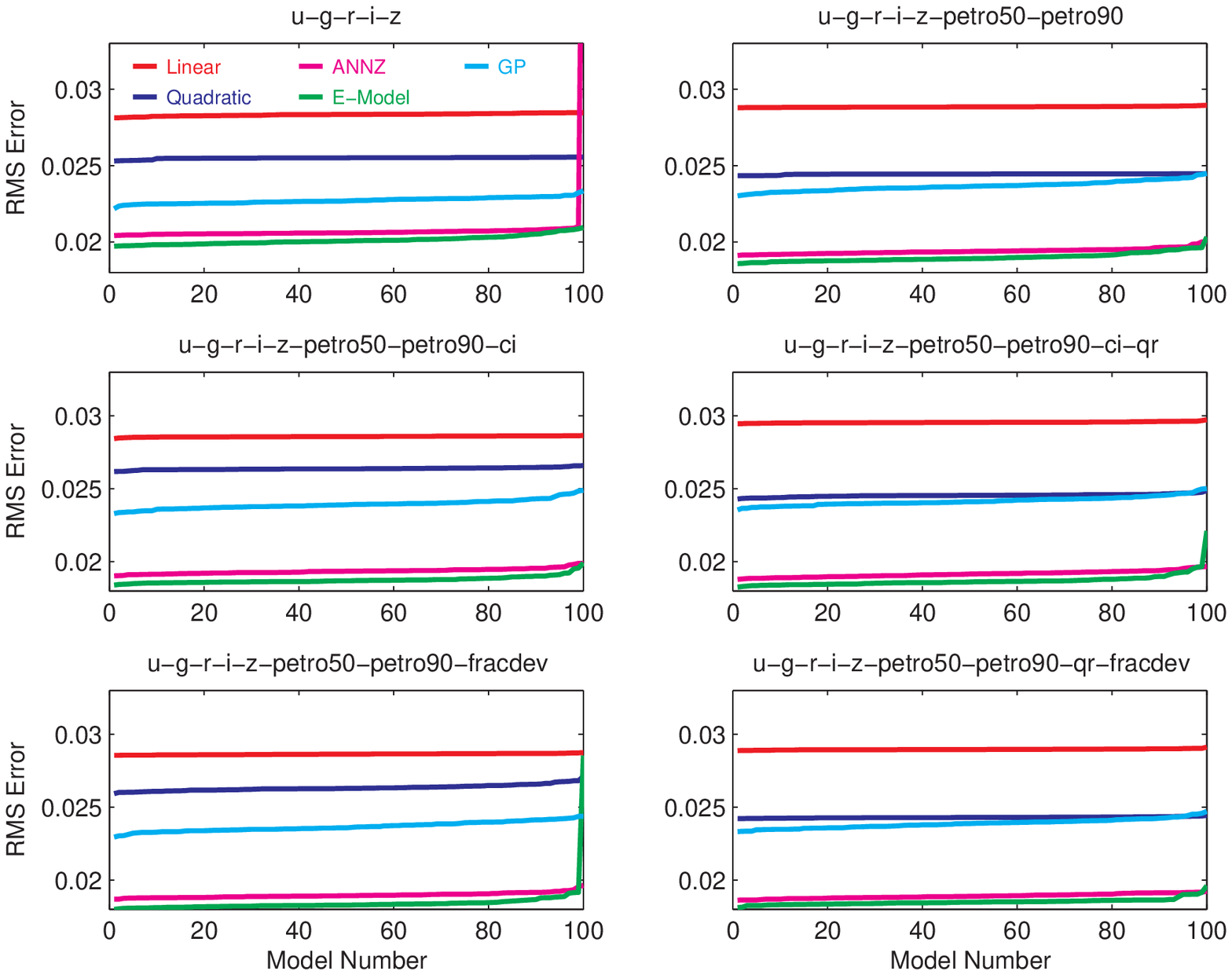}
\plotone{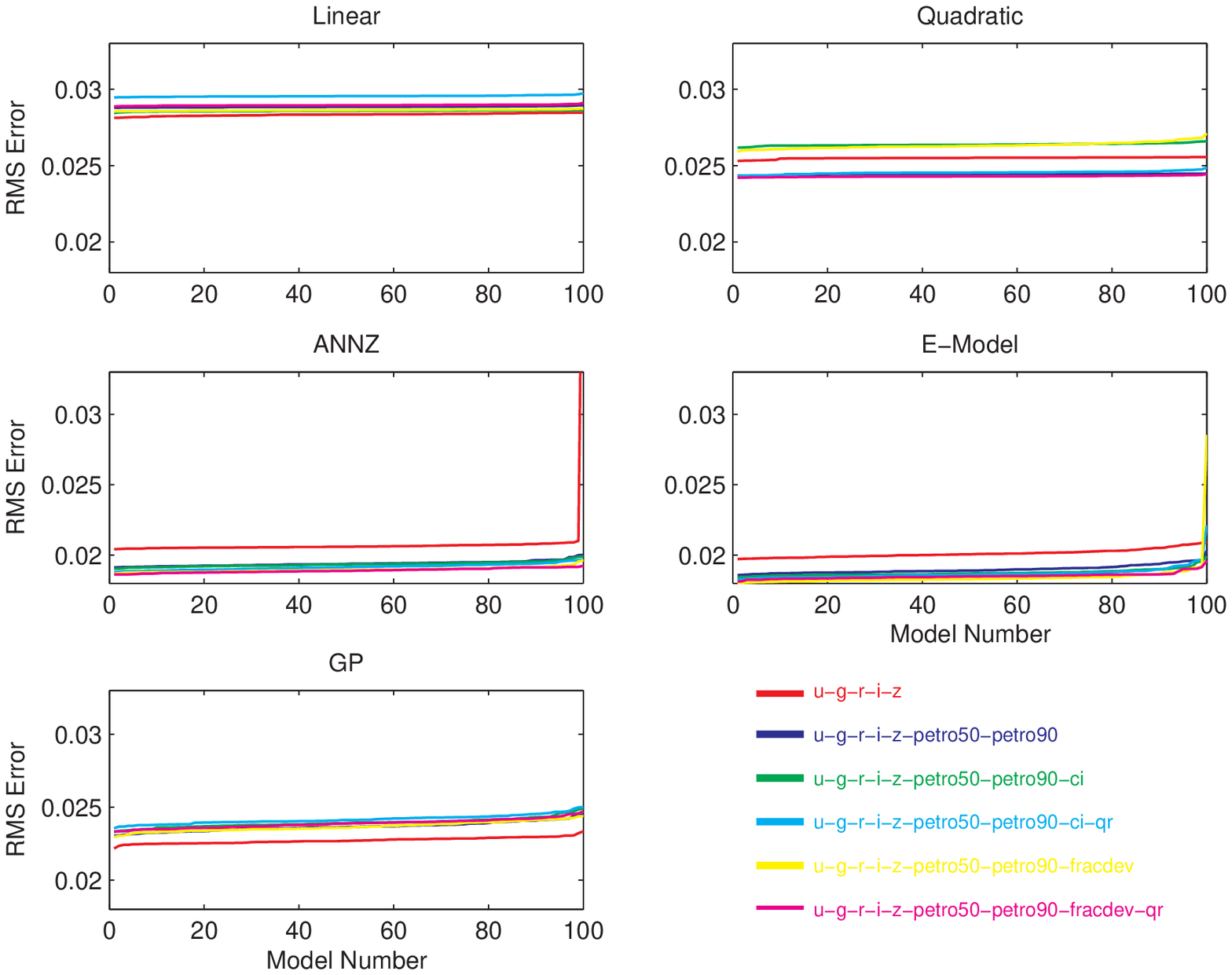}
\caption{(a)Top six plots containing the five training set methods
for each of the six inputs applied to the SDSS GOOD data sets
known as data set 1. (b) The bottom five plots are our training--set results
for each of the five training methods applied to the six different SDSS GOOD
inputs. } \label{fig:figure3} \vspace*{-.2in}
\end{figure}

\begin{figure}[dr3cresults]
\epsscale{0.8}
\plotone{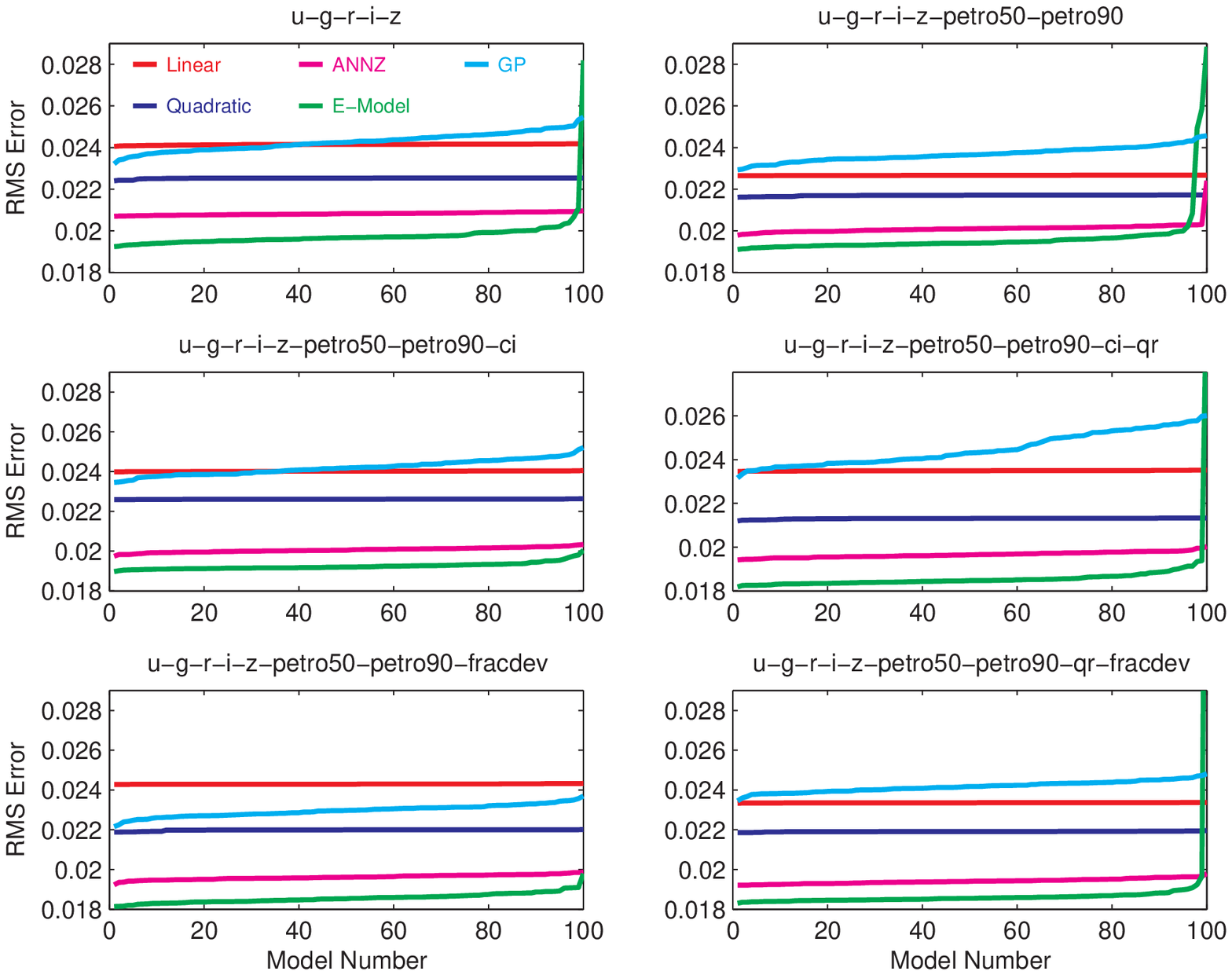}
\plotone{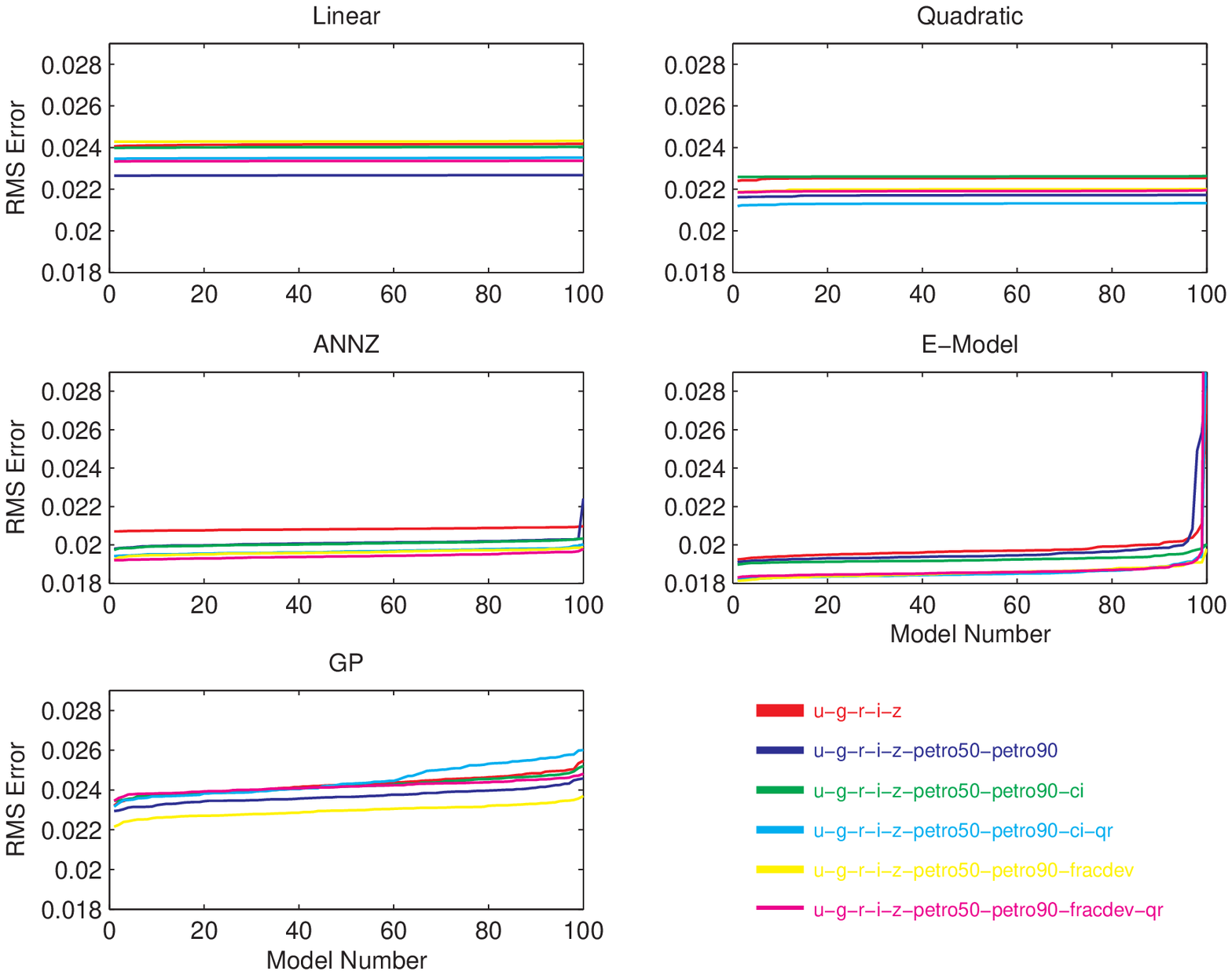}
\caption{(a) The top six plots contain the five training methods for
each of the six inputs applied to the SDSS GREAT data sets known as data set 2.
(b) The bottom five plots are our training--set results for each of the five
training methods applied to the six different SDSS GREAT inputs.}
\label{fig:figure4}
\vspace*{-.2in}
\end{figure}

\begin{figure}[galdr22mass]
\plotone{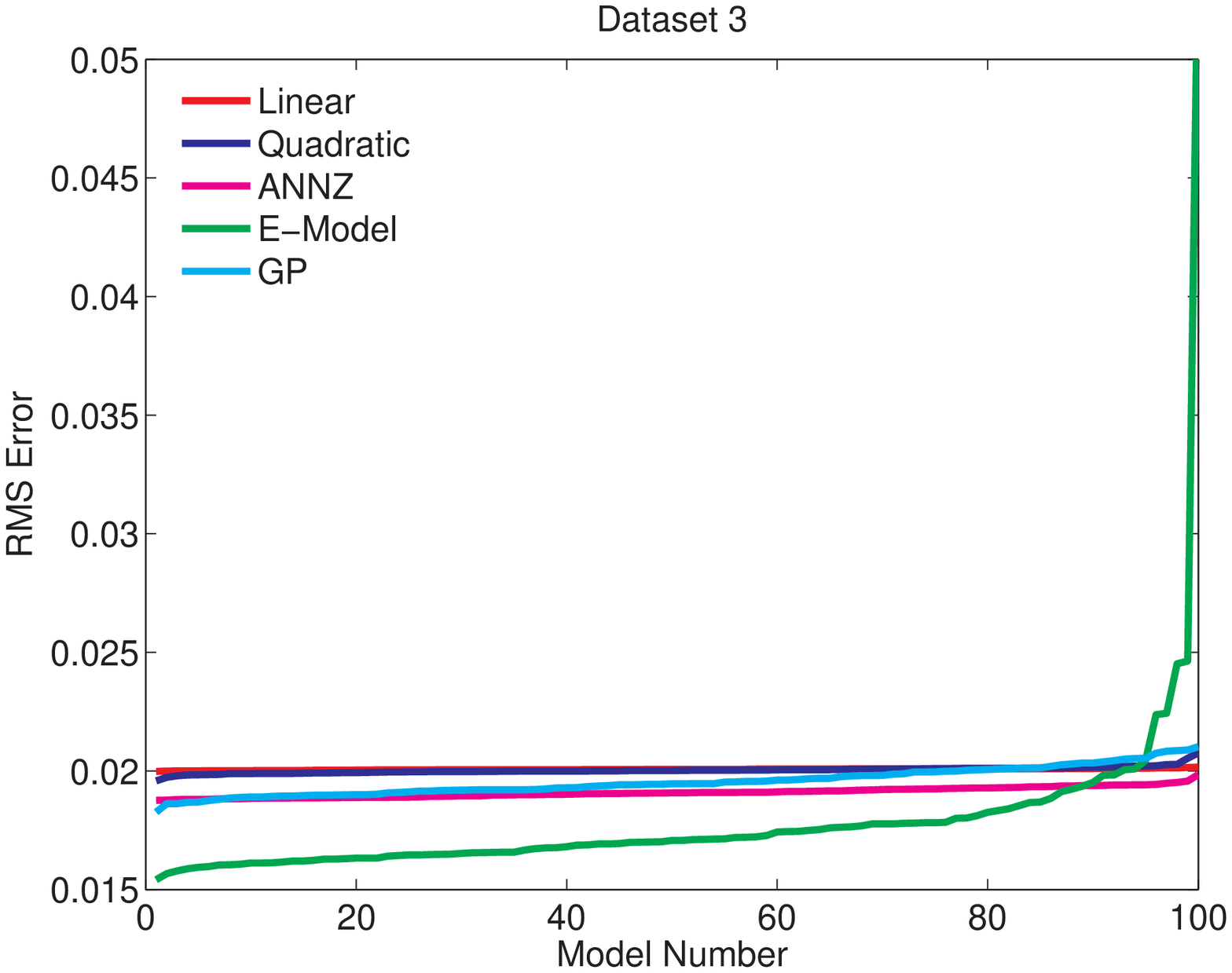}
\plotone{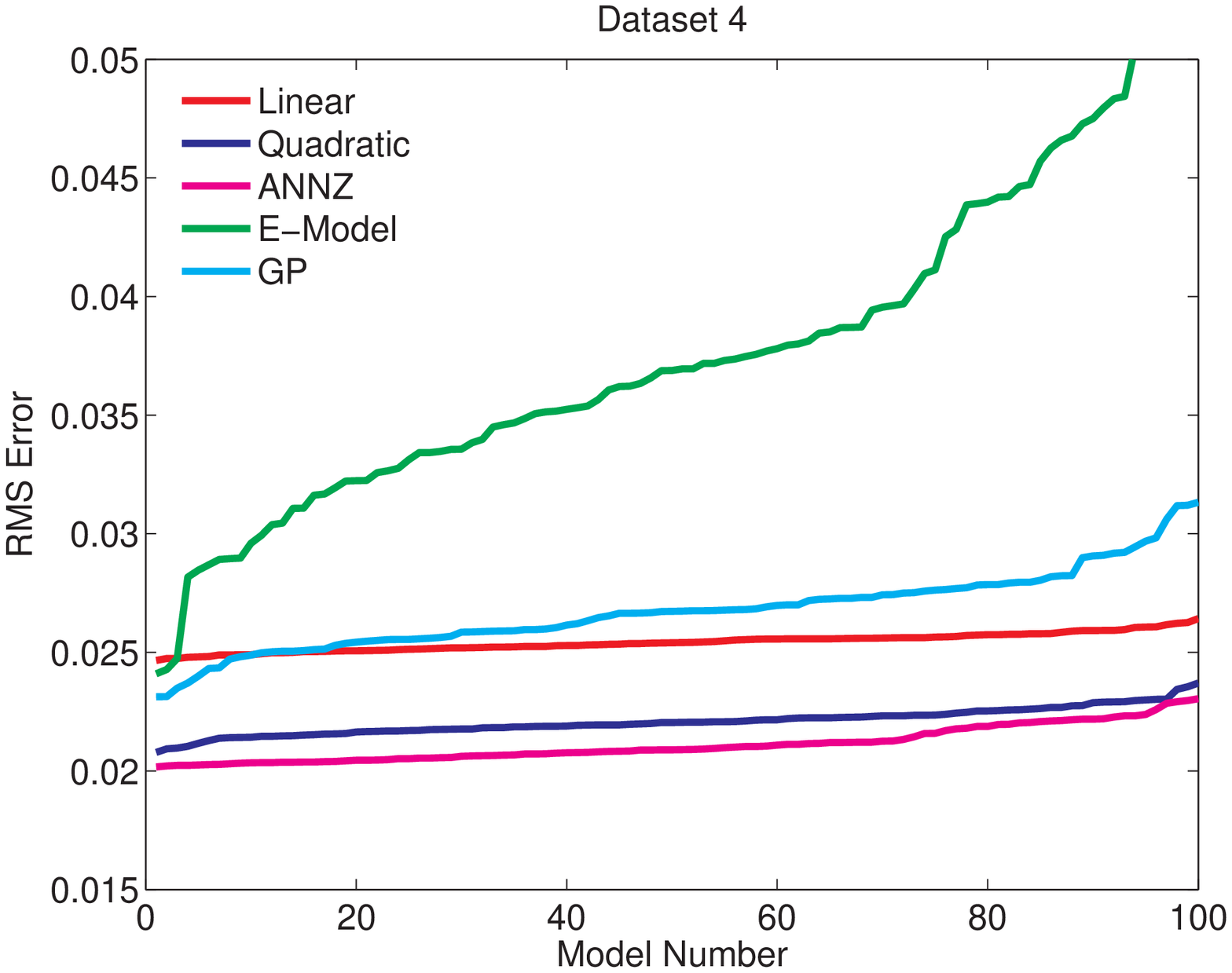}
\caption{(a) The top plot shows the five training methods applied to data
set 3. (b) The bottom  plot shows the five training methods applied to data
set 4.}
\label{fig:figure5}
\vspace*{-.2in}
\end{figure}

In Tables~\ref{tbl-4}--\ref{tbl-6} we report robust 90\% confidence intervals
around our 50\% RMS result for all of these methods from the bootstrap
resampling.  Figures~\ref{fig:figure3}, \ref{fig:figure4}
and \ref{fig:figure5} show the same information, albeit in a more detailed
graphical format.

Table~\ref{tbl-4} and Figure~\ref{fig:figure3} demonstrate our results on data
set 1.  The plots in Figure~\ref{fig:figure3} clearly demonstrate that the ANNz
and E--model neural network methods are superior in their accuracy over nearly
all bootstrap samples (labeled ``model number" in Figures 3-5) no matter which
input quantities are used.  The linear and quadratic fits fair the worse
as is expected. The Gaussian process model is usually found in between.
However, it must be remembered that only $\sim$1000 sample points are used
for training in each case and therefore it is possible that it is not sampling
all of the possible redshift--color space. Nonetheless it does an
excellent job given the small data samples used in comparison to the
other methods. It is also clear that the inputs used reproduce
very similar results once one goes beyond the five--band magnitudes of
the SDSS and quantities like the Petrosian concentration index or the
Stokes measure of ellipticity are used. The best method, our ensemble model,
regularly reproduces RMS values of less than 0.019 no matter the confidence
level (or bootstrap sample) used.

Table~\ref{tbl-5} and Figure~\ref{fig:figure4} for data set 2
give results very similar to those of data set 1 just discussed.
Lower RMS errors are obtained than that of the GOOD quality data, but
there is more variation in the confidence intervals evidenced by increasing
slope as a function of bootstrap sample in Figure~\ref{fig:figure4}.
As with data set 1, the RMS error results are lower but similar
when the five--band SDSS magnitudes are supplemented with quantities 
such as the Petrosian radii or the FracDev measurement.

While data set 2 does on occasion have slightly better RMS errors than data set
1, in general there is little difference in the use of higher quality photometry
and we would not recommend the use of the higher quality photometry of
data set 2 as described herein in general.

Table~\ref{tbl-6} and Figure~\ref{fig:figure5} are the results of using
data sets 3 and 4.  Figure~\ref{fig:figure5}b for data set 4 (which has better
photometric quality) shows again an increase in the variability of
the RMS error as a function of bootstrap sample larger than that of
the GOOD sample from data set 3 in Figure~\ref{fig:figure5}a.
In general Figure~\ref{fig:figure5}b
with the better SDSS photometry of data set 4 has RMS
errors either the same or worse than those from the SDSS only
data sets 1 and 2 in Figure~\ref{fig:figure3}
and~\ref{fig:figure4}. The variability in the RMS error as a
function of bootstrap and the generally large RMS errors leads one
to believe that the sample size is too small to train on. Given that
there are only 326 objects in data set 4 this should not be too
surprising. The apparent ability of the quadratic regression to do
so well might point one to possible over--fitting of the data.

However, in Figure~\ref{fig:figure5}a the story for
data set 3 is very different. Here the variability is much less a function of
bootstrap, the RMS errors are generally quite low, and the prediction abilities
of the different methods are consistent with those observed in the SDSS data
sets 1 and 2 found in Figures~\ref{fig:figure3} and \ref{fig:figure4}
The ensemble model once again surpasses all other methods for 95\% of
the bootstrap samples followed closely by the Gaussian processes and
ANNz methods. Here one can see that the Gaussian process method
is more competitive as it is likely to be sampling all possible templates
of the 3095 input galaxies even with only 1000 samples per run.

\begin{figure}[dr3cp5p9fq]
\plotone{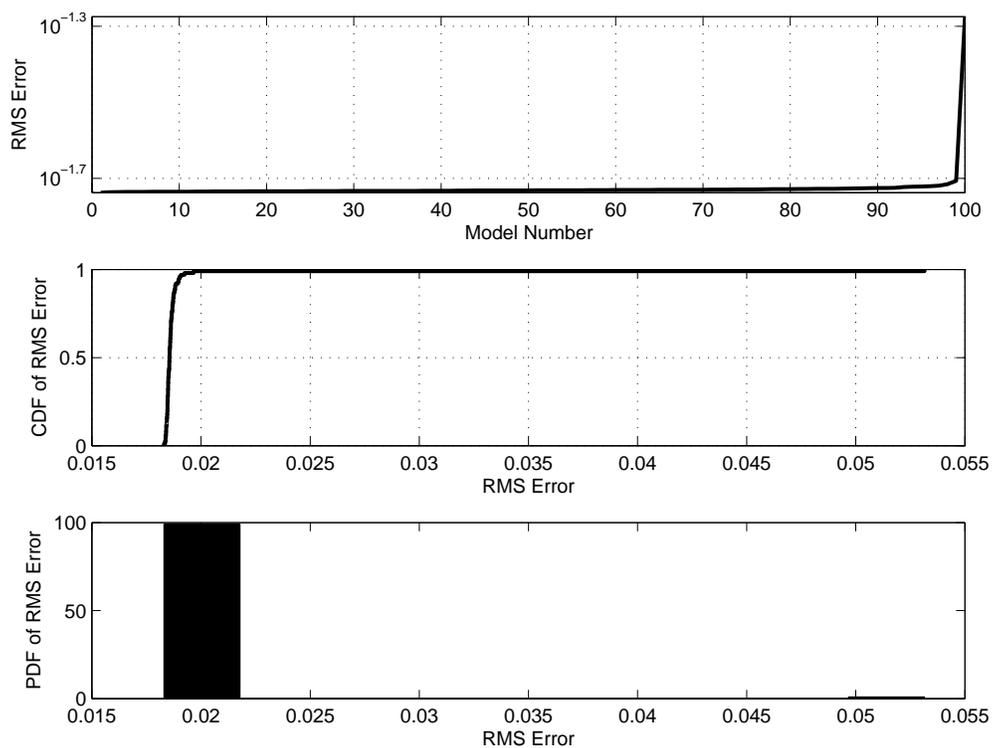}
\caption{The top panel of this figure shows the distribution of
errors for 100 neural networks on the GREAT E--Model
ugriz-petro50-petro90-qr-fracdev, data set 2 (see Table~\ref{tbl-5}). The middle
panel shows the empirical cumulative distribution function for the
RMS errors for the 100 models shown in the top panel. The bottom
panel shows the probability distribution function of the RMS error.
See {\S} 5 for more details.  } \label{fig:figure6}
\vspace*{-.2in}
\end{figure}

\begin{figure}[galexdr2c]
\plotone{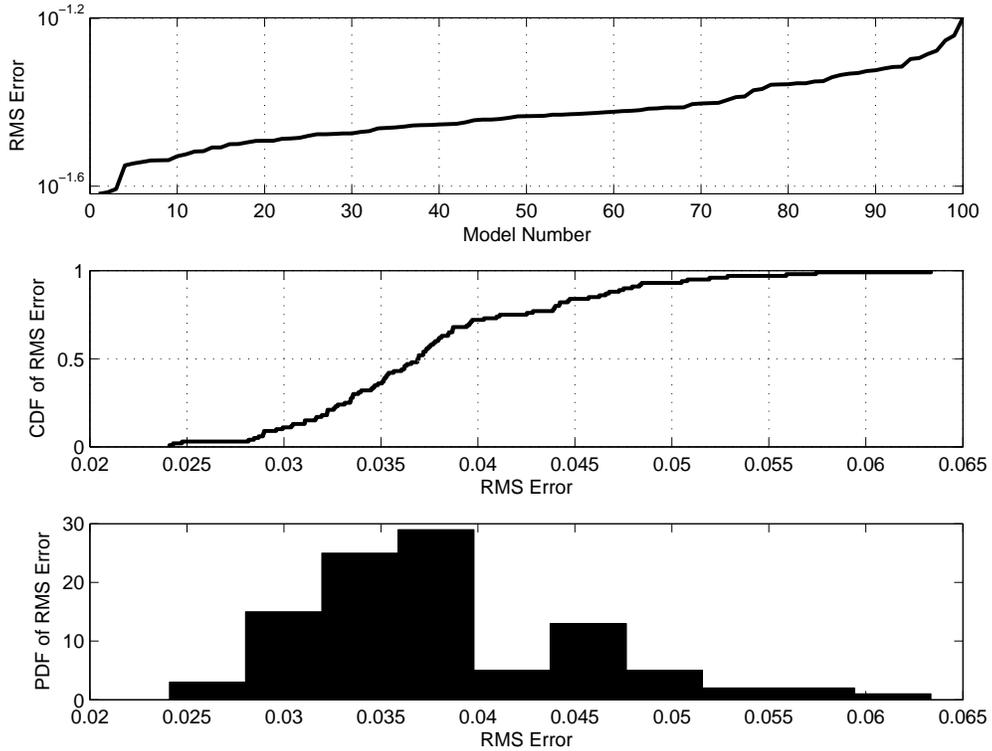}
\caption{The top panel of this figure shows the distribution of
errors for 100 neural networks on the GREAT E--Model nuv-fuv-ugriz-jhk, data set 4
(see Table~\ref{tbl-6}).  The middle panel shows the empirical cumulative
distribution function for the RMS errors for the 100 models shown in
the top panel. The bottom panel shows the probability distribution function of the RMS error.  See {\S} 5 for more details.} \label{fig:figure7}
\vspace*{-.2in}
\end{figure}

In order to show the effects of sampling and local minima for the ensemble
model on the quality of redshift predictions we show a set of 100 neural
networks and show their final RMS errors in
Figures~\ref{fig:figure6} and~\ref{fig:figure7}.  Each neural
network is built by drawing a sample from the training set with
replacement and then performing the gradient descent maximization
process described earlier.  We train until the model converges,
which is defined as the gradient--descent iteration at which the
magnitude of the gradient drops below a preset threshold.  This
model corresponds to one point on the top panel of
Figures~\ref{fig:figure6} and~\ref{fig:figure7}.

The middle panel of Figure~\ref{fig:figure7} shows the cumulative
distribution function for the errors shown in the top panel.  The
x-axis is the RMS error ($e_0$), and the y-axis is P(RMS $< e_0$).
The plot indicates that about 70\% of the models we generated have
an RMS error less than 0.1.  This plot also indicates that reporting
the minimum observed RMS value, which is done throughout the
literature on this topic \citep[e.g. ANNz]{Collister04} is
misleading. For the models computed for this empirical cumulative
distribution function, the quantity P(RMS $< e_0$) rapidly vanishes
as $e_0 \rightarrow 0.04$.  This implies that such models are not
only highly unlikely, but also highly non--robust.

In order to contrast this distribution with the empirical
distributions observed on other data sets, we chose to show
Figure~\ref{fig:figure6}.  This figure, unlike the previous
figure discussed, shows that the variation imposed by the optimization
procedure, combined with the variations in the data set, have a
relatively small effect on the quality of predictions:  nearly 70\%
of the models have a very low error rate, with the distribution
rapidly increasing after that.  Note that the empirical cumulative
distribution function shown in the bottom panel rises sharply at the
onset of the curve.  This indicates that 70\% of the models have an
error less than about 0.025.  Again, this variation and apparent
combined stability of the data set and optimization procedure would
be entirely lost if only the minimum value of the distribution was
reported.

\begin{figure}[specphot]
\centering
\includegraphics{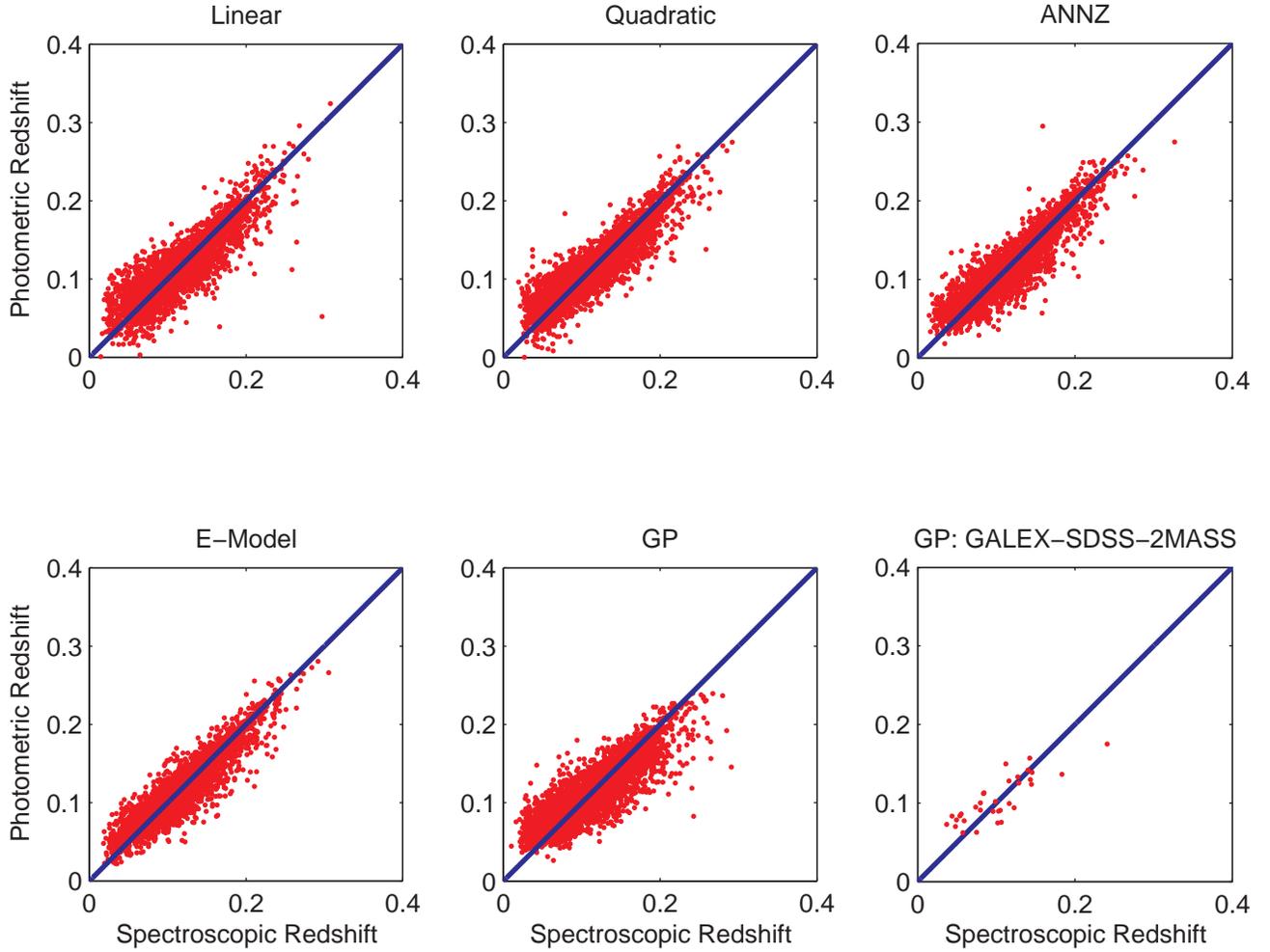}
\caption{Spectroscopic redshift is plotted versus calculated photometric
redshift for the GREAT ugriz-petro50-petro90-ci-qr data set 2 with 5 algorithms
while the 6th plot uses the Gaussian process model for the nuv-fuv-ugriz-jhk
GREAT data set 4. See Table~\ref{tbl-4} for details}
\label{fig:figure8}
\vspace*{-.2in}
\end{figure}

For comparison in Figure~\ref{fig:figure8} one can see the known spectroscopic
redshift plotted against the calculated photometric redshift from the test data
for our five algorithms used against the ugriz-petro50-petro90-ci-qr GREAT data
set (part of data set 2) as presented in Table~\ref{tbl-5}.
Note that the Gaussian process plot (bottom middle panel) has a larger
number of points, which is due to the smaller
training set and larger testing sets used in this algorithm.  The plot in
the bottom right--hand corner of Figure~\ref{fig:figure8} contains the
Gaussian process model results against the GREAT nuv-fuv-ugriz-jhk data set 4
as presented in Table~\ref{tbl-6}.

%\begin{figure}[dr3dresultsa]
%\subfigure[Six plots containing the five training set methods
%for each of the six inputs applied to the SDSS GOOD data sets
%known as data set 1.]{\label{fig:figure3a}\includegraphics{f3a.eps}}
%\newpage
%\subfigure[The five plots are our training--set results for each of the
%five training methods applied to the six different SDSS GOOD
%inputs.]{\label{fig:figure3b}\includegraphics{f3b.eps}}
%\caption{this is a test}
%\end{figure}

\begin{table}
\scriptsize
\begin{center}
\caption{Different Photometric Redshift Techniques and Accuracies. \label{tbl-3}}
\begin{tabular}{lclll}
\hline \hline
Method Name               &$\sigma_{RMS}$ & Data set\tablenotemark{1} & Inputs\tablenotemark{2} & Source \\
\hline
CWW                       & 0.0666        & SDSS-EDR &ugriz         & \cite{Csabai03} \\
Bruzual-Charlot           & 0.0552        & SDSS-EDR &ugriz         & \cite{Csabai03} \\
ClassX                    & 0.0340        & SDSS-DR2 &ugriz         & \cite{Suchkov05} \\
Polynomial                & 0.0318        & SDSS-EDR &ugriz         & \cite{Csabai03} \\
Support Vector Machine    & 0.0270        & SDSS-DR2 &ugriz         & \cite{Wadadekar05} \\
Kd-tree                   & 0.0254        & SDSS-EDR &ugriz         & \cite{Csabai03} \\
Support Vector Machine    & 0.0230        & SDSS-DR2 &ugriz+r50+r90 & \cite{Wadadekar05} \\
Artificial Neural Network & 0.0229        & SDSS-DR1 &ugriz         & \cite{Collister04} \\
Artificial Neural Network & 0.022-0.024   & SDSS-DR1 &A             & \cite{Vanzella04} \\
Artificial Neural Network & 0.0200-0.025  & SDSS-EDR &B             & \cite{Tagliaferri03} \\
Artificial Neural Network & 0.0200-0.026  & SDSS-EDR &C             & \cite{Ball04} \\
Polynomial                & 0.025          & SDSS-DR1,GALEX&ugriz+nuv& \cite{Budavari05} \\
\hline \hline
\end{tabular}
\tablenotetext{1}{SDSS-EDR Early Data Release \citep{Stoughton02},
SDSS-DR1 Data Release 1 \citep{Abazajian03},
SDSS-DR2 Data Release 2 \citep{Abazajian04}}
\tablenotetext{2}{ugriz=5 SDSS magnitudes, r50=Petrosian 50\%
light radius in r band, r90=Petrosian 90\% light radius in r band,
nuv=Near-Ultraviolet GALEX bandpass.
For A see \cite{Vanzella04}, for B see \cite{Tagliaferri03} and
for C see \cite{Ball04} for a list of the large variety of inputs used
in each of these publications.}
\end{center}
\end{table}

\begin{deluxetable}{l|ccc|ccc|ccc|ccc|ccc}
\tablecolumns{16}
\rotate
\tablewidth{0pc}
\tablecaption{Photometric Redshift prediction RMS errors with confidence levels for Dataset 1, 202,297 objects\label{tbl-4}}
\tabletypesize{\scriptsize}
\tablehead{
\colhead{Input-parameters\tablenotemark{1}}&
& \colhead{Linear}&&&\colhead{Quadratic}&&&\colhead{ANNz}&&&\colhead{E-Model}&&&\colhead{GP}&\\
& 
\colhead{(50\%)}&\colhead{(10\%)}&\colhead{(90\%)}&
\colhead{(50\%)}&\colhead{(10\%)}&\colhead{(90\%)}&
\colhead{(50\%)}&\colhead{(10\%)}&\colhead{(90\%)}&
\colhead{(50\%)}&\colhead{(10\%)}&\colhead{(90\%)}&
\colhead{(50\%)}&\colhead{(10\%)}&\colhead{(90\%)}
}
\startdata
ugriz              &0.0283&0.0282&0.0284&0.0255&0.0255&0.0255&0.0206&0.0205&0.0208&0.0201&0.0198&0.0205&0.0227&0.0225&0.0230\\
ugriz+r50+r90      &0.0288&0.0288&0.0289&0.0245&0.0244&0.0245&0.0194&0.0192&0.0196&0.0189&0.0187&0.0194&0.0236&0.0233&0.0241\\
ugriz+r50+r90+CI   &0.0286&0.0285&0.0286&0.0264&0.0263&0.0265&0.0194&0.0191&0.0195&0.0187&0.0185&0.0190&0.0239&0.0236&0.0243\\
ugriz+r50+r90+CI+QR&0.0296&0.0295&0.0296&0.0245&0.0244&0.0246&0.0192&0.0189&0.0194&0.0186&0.0184&0.0190&0.0241&0.0238&0.0245\\
ugriz+r50+r90+FD   &0.0286&0.0286&0.0287&0.0263&0.0261&0.0266&0.0189&0.0188&0.0192&0.0183&0.0181&0.0187&0.0236&0.0233&0.0241\\
ugriz+r50+r90+FD+QR&0.0290&0.0289&0.0290&0.0243&0.0242&0.0243&0.0189&0.0187&0.0191&0.0185&0.0183&0.0186&0.0239&0.0235&0.0242\\
\enddata
\tablenotetext{1}{ugriz=5 SDSS magnitudes, r50=Petrosian 50\% light radius in
r band, r90=Petrosian 90\% light radius in r band, CI=Petrosian Inverse
Concentration Index, FD=FracDev value, QR=Stokes value.  See {\S} 3.6 for more
details.}
\end{deluxetable}

\begin{deluxetable}{l|ccc|ccc|ccc|ccc|ccc}
\tablecolumns{16}
\rotate
\tablewidth{0pc}
\tablecaption{Photometric Redshift prediction RMS errors with confidence levels for Dataset 2, 33,328 objects\label{tbl-5}}
\tabletypesize{\scriptsize}
\tablehead{
\colhead{Input-parameters\tablenotemark{1}}&
& \colhead{Linear}&&&\colhead{Quadratic}&&&\colhead{ANNz}&&&\colhead{E-Model}&&&\colhead{GP}&\\
&
\colhead{(50\%)}&\colhead{(10\%)}&\colhead{(90\%)}&
\colhead{(50\%)}&\colhead{(10\%)}&\colhead{(90\%)}&
\colhead{(50\%)}&\colhead{(10\%)}&\colhead{(90\%)}&
\colhead{(50\%)}&\colhead{(10\%)}&\colhead{(90\%)}&
\colhead{(50\%)}&\colhead{(10\%)}&\colhead{(90\%)}
}
\startdata
ugriz              &0.0242&0.0241&0.0242&0.0225&0.0225&0.0225&0.0208&0.0207&0.0209&0.0197&0.0194&0.0200&0.0243&0.0237&0.0248\\
ugriz+r50+r90      &0.0227&0.0227&0.0227&0.0217&0.0216&0.0217&0.0201&0.0199&0.0202&0.0194&0.0192&0.0198&0.0237&0.0232&0.0241\\
ugriz+r50+r90+CI   &0.0240&0.0240&0.0240&0.0226&0.0226&0.0226&0.0200&0.0199&0.0202&0.0192&0.0191&0.0194&0.0242&0.0238&0.0247\\
ugriz+r50+r90+CI+QR&0.0235&0.0235&0.0235&0.0213&0.0213&0.0213&0.0197&0.0195&0.0198&0.0185&0.0183&0.0189&0.0243&0.0237&0.0255\\
ugriz+r50+r90+FD   &0.0243&0.0243&0.0243&0.0220&0.0219&0.0220&0.0196&0.0195&0.0198&0.0185&0.0183&0.0189&0.0230&0.0226&0.0233\\
ugriz+r50+r90+FD+QR&0.0234&0.0233&0.0234&0.0220&0.0219&0.0220&0.0194&0.0193&0.0196&0.0185&0.0184&0.0188&0.0242&0.0238&0.0245\\
\enddata
\tablenotetext{1}{ugriz=5 SDSS magnitudes, r50=Petrosian 50\% light radius
in r band, r90=Petrosian 90\% light radius in r band, CI=Petrosian Inverse
Concentration Index, FD=FracDev value, QR=Stokes value. See {\S 3.6} for more
details.}
\end{deluxetable}

\begin{deluxetable}{l|ccc|ccc|ccc|ccc|ccc}
\tablecolumns{19}
\rotate
\tablewidth{0pc}
\tablecaption{Photometric Redshift prediction RMS errors with confidence levels for Datasets 3 and 4.\label{tbl-6}}
\tabletypesize{\scriptsize}
\tablehead{
\colhead{Input-parameters\tablenotemark{1}}&&\colhead{Linear}&&&\colhead{Quadratic}&&&\colhead{ANNz}&&&\colhead{E-Model}&&&\colhead{GP}&\\
&\colhead{(50\%)}&\colhead{(10\%)}&\colhead{(90\%)}&
\colhead{(50\%)}&\colhead{(10\%)}&\colhead{(90\%)}&
\colhead{(50\%)}&\colhead{(10\%)}&\colhead{(90\%)}&
\colhead{(50\%)}&\colhead{(10\%)}&\colhead{(90\%)}&
\colhead{(50\%)}&\colhead{(10\%)}&\colhead{(90\%)}}
\startdata
nuv+fuv+ugriz+jhk\tablenotemark{2}&0.0201&0.0200&0.0201&0.0200&0.0199&0.0202&0.0191&0.0188&0.0194&0.0171&0.0161&0.0195&0.0195&0.0189&0.0203\\
nuv+fuv+ugriz+jhk\tablenotemark{3}&0.0254&0.0249&0.0259&0.0220&0.0214&0.0229&0.0209&0.0204&0.0222&0.0369&0.0296&0.0475&0.0267&0.0249&0.0291\\
\enddata
\tablenotetext{1}{ugriz=5 SDSS magnitudes, nuv=GALEX NUV magnitude, fuv=GALEX
FUV magnitude, jhk=2MASS jkh magnitudes. See {\S} 3.6 for more details.}
\tablenotetext{2}{Dataset 3: 3095 combined catalog objects}
\tablenotetext{3}{Dataset 4: 326 combined catalog objects}
\end{deluxetable}

\section{CONCLUSION}

We have shown that photometric redshift accuracy of SDSS photometric
data can be improved over that of previous attempts through
a careful choice of additional photometric pipeline outputs that
are related to angular size and morphology. Adding additional
bandpasses from the ultraviolet (GALEX) and infrared (2MASS) can be even
more helpful, but the current sample sizes are too small
to be useful for large--scale structure studies.

We have also shown that there is little difference in the use of the
higher quality SDSS photometry as defined herein. Hence we would not
recommend its use because it decreases the sample size markedly and
does not decrease the RMS errors in the photometric redshift
prediction.

We wish to stress that when using a neural network model for studies of
photometric redshifts care must be taken when reporting the results
of such models. There is a tendency in the astronomical literature to
report only the best--fit model, which is often unlikely to be the
one used to calculate the final photometric redshift estimates.

The effects of local minima on prediction have also been discussed in some
detail and we describe the way in which an ensemble of neural networks
can reduce the problem.

We have also discussed the result of using Gaussian processes for regression,
which avoids many of the local minima problems that occur with neural networks.
One of the great strengths of Gaussian processes as used herein
is the ability to use small training sets, which may be
helpful in high--redshift studies where very small numbers of
measured redshifts are available.

Finally, it should be noted that the TS methods described herein are only useful
in a limited set of circumstances. In this work the SDSS MGS has been utilized
since it is considered a complete photometric and spectroscopic survey in the
sense that the magnitude limit of the survey is well understood, a broad range
of colors are measured, and accurate redshifts obtained. It would be folly to
attempt to use TS methods in a situation where these are poorly defined. For
example, to simply apply TS methods to the entire SDSS galaxy photometric and
redshift catalog without taking into account the limitations in the quantity
and quality of photometry and redshifts would likely give one results that
could not be quantified properly and give misleading conclusions. As well,
it has been stressed that TS--methods have not been widely used in z$>$1 surveys
because thus far a complete sample of redshifts over the observed colors and
magnitudes of the galaxies of interest have not been measured. This will change
as larger telescopes with more sensitive detectors appear, but TS methods will
not be useful for those situations where insufficient numbers
of redshifts, colors and magnitudes exist to cover the required spaces.

%%%%%%%%%%%%%%%%%%%%%%%%%%%%%%%%%%%%%%%%%%%%%%%%%%%%%%%%%%%%%%%%%%%%%%%

\acknowledgements

M. J. W. acknowledges useful discussions with \v{Z}eljko Ivezi\'{c} on
the SDSS photometric quality flags, and Michael Blanton related to
the NYU-VAGC catalog, which helped in understanding many aspects of
the SDSS. Creon Levit and Paul Gazis also provided useful information
related to neural networks, and Jeff Scargle gave critical input by
the careful reading of our manuscript. M. J. W. also acknowledges
Alex Szalay, Ani Thakar, Maria SanSebastien, and Jim Gray for their
help in using the SDSS skyserver query interface and in obtaining a
local copy of the SDSS DR2. M. J. W. acknowledges funding received from
the NASA Applied Information Systems Research Program. A. N. S.
acknowledges funding received from the NASA Intelligent Systems Program,
Intelligent Data Understanding Element, and valuable
discussions with William Macready.

The authors acknowledge support from the NASA Ames Research Center
Director's Discretionary Fund.

Funding for the SDSS has been provided by
the Alfred P. Sloan Foundation, the Participating Institutions, the National
Aeronautics and Space Administration, the National Science Foundation,
the U.S. Department of Energy, the Japanese Monbukagakusho, and the Max
Planck Society. The SDSS Web site is http://www.sdss.org/.

The SDSS is managed by the Astrophysical Research Consortium for
the Participating Institutions. The Participating Institutions are the
University of Chicago, Fermilab, the Institute for Advanced Study, the
Japan Participation Group, the Johns Hopkins University, Los Alamos National
Laboratory, the Max Planck Institute for Astronomy, the
Max Planck Institute for Astrophysics, New Mexico State University, the
University of Pittsburgh, Princeton University, the United States Naval
Observatory, and the University of Washington.

This publication makes use of data products from the Two Micron All Sky Survey,
which is a joint project of the University of Massachusetts and the Infrared
Processing and Analysis Center/California Institute of Technology, funded by
the National Aeronautics and Space Administration and the National Science
Foundation.

This research has made use of NASA's Astrophysics Data System bibliographic
services.

%%%%%%%%%%%%%%%%%%%%%%%%%%%%%%%%%%%%%%%%%%%%%%%%%%%%%%%%%%%%%%%%%%%%%%%

\appendix

\section{SDSS QUERIES}

Below are the queries used against the SDSS DR2 and DR3 databases
to obtain the data used throughout this paper.

\noindent Query used to obtain data set 1:

\noindent Select p.ObjID, p.ra, p.dec,\newline p.dered\_u,
p.dered\_g, p.dered\_r, p.dered\_i, p.dered\_z,\newline
p.petroR50\_r, p.petroR90\_r, p.fracDeV\_r, p.q\_r,\newline
p.Err\_u, p.Err\_g, p.Err\_r, p.Err\_i, p.Err\_z,\newline
p.petroR50Err\_r, p.petroR90Err\_r, p.qErr\_r,\newline s.z, s.zErr,
s.zConf\newline into mydb.dr3cfracdpetq from  SpecOBJall s,
PhotoObjall p\newline WHERE s.specobjid=p.specobjid\newline and
s.zConf$>$0.95\newline and (p.primtarget \& 0x00000040 $>$
0)\newline and ( ((flags \& 0x8) = 0) and ((flags \& 0x2) = 0) and
((flags \& 0x40000) = 0))\newline

\noindent Query used to obtain data set 2:

\noindent Select p.ObjID, p.ra, p.dec,\newline
p.dered\_u, p.dered\_g, p.dered\_r, p.dered\_i, p.dered\_z,\newline
p.petroR50\_r, p.petroR90\_r, p.fracDeV\_r, p.q\_r,\newline
p.Err\_u, p.Err\_g, p.Err\_r, p.Err\_i, p.Err\_z,\newline
p.petroR50Err\_r, p.petroR90Err\_r, p.qErr\_r,\newline
s.z, s.zErr, s.zConf\newline
into mydb.dr3cfracdpetq from  SpecOBJall s, PhotoObjall p\newline
WHERE s.specobjid=p.specobjid\newline
and s.zConf$>$0.95\newline
and (p.primtarget \& 0x00000040 $>$ 0)\newline
and ( ((flags \& 0x8) = 0) and ((flags \& 0x2) = 0) and ((flags \& 0x40000) = 0)\newline
and   ((flags \& 0x10) =0) and ((flags \& 0x1000)=0) and ((flags \& 0x20000) = 0) )\newline

\noindent Query used to obtain data set 3:

\noindent Select p.objID, p.ra, p.dec,\newline
g.NUV\_MAG, g.NUV\_MAGERR, g.FUV\_MAG, g.FUV\_MAGERR,\newline
p.u, p.Err\_u, p.g, p.Err\_g, p.r, p.Err\_r, p.i, p.Err\_i, p.z, p.Err\_z,\newline
t.j\_m\_k20fe, t.j\_msig\_k20fe, t.h\_m\_k20fe, t.h\_msig\_k20fe, t.k\_m\_k20fe, t.k\_msig\_k20fe, \newline
s.z, s.zErr, s.zConf\newline
FROM TWOMASS.dbo.xsc t, BESTDR2.dbo.PhotoObjAll p, GALEXDRONE.dbo.nuvfuv g, BESTDR2.dbo.SpecOBJall s\newline
WHERE s.specobjid=p.specobjid\newline
and s.zConf$>$0.95 and s.zWarning=0\newline
and g.NUV\_MAG$>$-99 and g.FUV\_MAG$>$-99\newline
and t.cc\_flg='0'\newline
and (p.primtarget \& 0x00000040 $>$ 0)\newline
and ((flags \& 0x8)=0) and ((flags \& 0x2)=0) and ((flags \& 0x40000)=0)\newline
and p.objid=BESTDR2.dbo.fgetnearestobjideq(t.ra,t.dec,0.08333)\newline
and p.objid=BESTDR2.dbo.fgetnearestobjideq(g.RA,g.DEC,0.08333)\newline
\newline

\noindent Query used to obtain data set 4:

\noindent Select p.objID, p.ra, p.dec,\newline
g.NUV\_MAG, g.NUV\_MAGERR, g.FUV\_MAG, g.FUV\_MAGERR,\newline
p.u, p.Err\_u, p.g, p.Err\_g, p.r, p.Err\_r, p.i, p.Err\_i, p.z, p.Err\_z,\newline
t.j\_m\_k20fe, t.j\_msig\_k20fe, t.h\_m\_k20fe, t.h\_msig\_k20fe, t.k\_m\_k20fe, t.k\_msig\_k20fe, \newline
s.z, s.zErr, s.zConf\newline
FROM TWOMASS.dbo.xsc t, BESTDR2.dbo.PhotoObjAll p, GALEXDRONE.dbo.nuvfuv g, BESTDR2.dbo.SpecOBJall s\newline
WHERE s.specobjid=p.specobjid\newline
and s.zConf$>$0.95 and s.zWarning=0\newline
and g.NUV\_MAG$>$-99 and g.FUV\_MAG$>$-99\newline
and t.cc\_flg='0'\newline
and (p.primtarget \& 0x00000040 $>$ 0)\newline
and ( ((flags \& 0x8)=0) and ((flags \& 0x2)=0) and ((flags \& 0x40000)=0)\newline
and   ((flags \& 0x10) =0) and ((flags \& 0x1000)=0) and ((flags \& 0x20000) = 0) )\newline
and p.objid=BESTDR2.dbo.fgetnearestobjideq(t.ra,t.dec,0.08333)\newline
and p.objid=BESTDR2.dbo.fgetnearestobjideq(g.RA,g.DEC,0.08333)\newline
\newline

\end{document}